# Transformable Biomimetic Liquid Metal Chameleon

Shuting Liang[1], Wei Rao[2], Kai Song[2], Jing Liu[1,2]∗

1. Department of Biomedical Engineering, School of Medicine, Tsinghua University,
Beijing 100084, China
2. Technical Institute of Physics and Chemistry, Chinese Academy of Sciences,
Beijing 100190, China
∗ E-mail: jliubme@tsinghua.edu.cn

**Abstract**

Liquid metal (LM) is of current core interest for a wide variety of newly emerging areas. However, the functional materials thus made so far by LM only could display a single silver-white appearance. Here in this study, the new conceptual colorful LM marbles working like transformable biomimetic "chameleons" were proposed and fabricated from LM droplets through encasing them with fluorescent nano-particles. We demonstrated that this unique LM marble can be manipulated into various stable magnificent appearances as one desires. And it can also splitt and merge among different colors. Such multifunctional LM "chameleon" is capable of responding to the outside electric-stimulus and realizing shape transformation and discoloration behaviors as well. Further more, the electric-stimuli has been disclosed to be an easy going way to trigger the release of nano/micro-particles from the LM. The present fluorescent biomimetic liquid metal chameleon is expected to offer important opportunities for diverse unconventional applications, especially in a wide variety of functional smart material and color changeable soft robot areas.

## 1. Introduction

Nature has created a wonderful world full of beautiful colors. As an interesting animal, the chameleon owns the remarkable ability to rapidly change its skin colors during the interactions like courtship, camouflage and male contests [1]. In engineering area, quiet a few of the bionic-robots have been fabricated to imitate the color-changing creatures in nature. Liquid metal (LM: EGaIn), has distinctive high flexibility, deformability, electrical conductivity, thermal conductivity, and nontoxicity [2, 3]. This enables its a multitude of unique virtues in a wide range of applications including the flexible bionic-robot [4-6], large-scale printed circuit board (PCB) [7-10], flexible electronics [11-15], heat sink [16], 3D metal printing [17, 18], sensors [19] or batteries [20], skin electronics [21-24], microfluidic channels [25-27], functional devices [28, 29], and medical field [30, 31] etc. For this reason, liquid metal has recently received increasing attentions spanning from diverse engineering disciplines to fundamental sciences.

At the present stage, the LM EGaIn has a high reflectivity as most of other metals own in the optical aspect. Thus it often presented a silver-white hue [32]. After utilizing the LM for the fabrication, the functional commodities usually got single appearances as well. From conceptual innovation aspect, here we envision that a big challenge in the field would come to the actualization of LM for full-color appearances which could be applied to a host of unusual practices.

In order to make the color changeable object, the LM marble was fast intensionally oxidized in an oxygen environment [33] and constructed as a thin $Ga_2O_3$ oxide film with a thickness of 0.5~2.5nm [34, 35]. The oxide shell has been described with an outstanding adhesion behavior and strongly adhered to whatever surfaces [15, 36, 37]. The theory of the adhesion process of the oxide film has been established [38]. According to the previous reports, the oxidation films of LM marbles coated with $WO_3$, $TiO_2$, $In_2O_3$, C nano-tube and semiconductor nano-material [33, 39] have a sea of sensitive electrochemical [40] and photochemical [41] properties. And some of these white LM marbles coated with hydrophobic coverings can even float on the water. Besides, a magnetic LM marble was acquired by covering an LM with nano-sized Fe particles [42], which has a large potential in the magnetic field-driven manipulation [43], or electrical switching field [44]. Consequently, the coating of the fluorescent grains on the superficies of the LM marbles became the key ingredient due to their unique functions.

Herein we described the primary instance of a spontaneous adhesion of stabilized fluorescent particulates on the superficies of LM droplets to achieve colorizing the LM droplet. The evidence suggested that this fluorescent LM marble was rather stable. And it can be merged and split between each other. This leads to the establishment of a new conceptual fluorescent biomimetic liquid metal chameleon for the first time. Further, under the electro stimuli, we revealed the transformation and discoloration behavior of such fluorescent LM marble which may suggest some important applications in the coming time.

## 2. Materials and Methods

The liquid metal (eutectic gallium indium, EGaIn) was made by high-purity Gallium and Indium metals (with purity of 99.9%) as raw materials. Gallium and Indium were weighted with a ratio of 75.5:24.5 in a line of the chemical compositions of $GaIn_{24.5}$ alloy. The weighted Gallium and Indium metals were mixed together in a beaker and heated to 50$^o$C until the metals were fused completely. Different kinds of commercial fluorescent powders were adopted as the coating materials. The composition of rare earth phosphors is: red phosphor: $YVO_4$: $Eu^{3+}$; yellow phosphor: $Y_2Al_5O_{12}$: $Ce^{3+}$; green phosphor: $(Ba,Si)_2SiO_4$: $Eu^{2+}$, and blue phosphor: $BaMgAl_{10}O_{17}$: $Eu^{3+}$. Commercial fluorescent powders and the copper electrodes were all purchased from Sinopharm Chemical Group Co., Ltd, and the purity of the two materials were all as high as 99.9%.

The result of fluorescent LM droplet was characterized through a field emission scanning electron microscope (QUANTA FEG 250). The surface of the fluorescence LM was obtained by the SEM images. The optical properties of the fluorescent LM droplets were analyzed by diffuse reflectance spectroscopy (UV-vis), which was performed using a Hitachi (U-3010) Spectrophotometer in the wavelength range of 300-780nm employs barium sulfate as a reference. Fluorescence spectra measurements were performed on a FLSP-920 fluorescence spectrophotometer. The fluorescence of marbles were measured at room temperature. The electric field is imposed by applying a DC between two inert copper rods as electrodes. The DC voltages are generated by a DC power supply (RIGOL, DG 1032Z 30-MHz Dual Channel, China) while the ramping voltages are obtained by a lab view controlled source meter with data acquisition function. High-speed videos of the autonomous locomotion are captured using a camera (Canon XF-305, Japan), and the transient images are extracted from these videos.

## 3. Results and Discussion

Two attempts have been devoted for fabricating the fluorescent LM droplets: (1) As illustrated in **Figure 1(a)**, agitating the LM with rare-earth phosphors in a vessel. The fluorescent powder of 2g was mixed with the LM of 20g in a beaker, which was maintained under continuous stirring at 60$^o$C. The growth of LM oxide film (consists mainly of $Ga_2O_3$) was accelerated in the atmosphere [33, 45]. The naturally oxidized LM has a viscoelastic oxidative tegument. The fluorescent powders were substantially adhered to the LM droplet's oxidation tegument after being fully stirred for more than 2h, and the adhesion reason would be discussed in the following result section; (2) Alternative method of overlaying the LM marble was to roll a LM droplet on a fluorescent powder bed, which was overspread with the fluorescent powders. The fluorescence properties prepared by the second method were not as perfect as the first one, which may be due to an insufficient stirring.

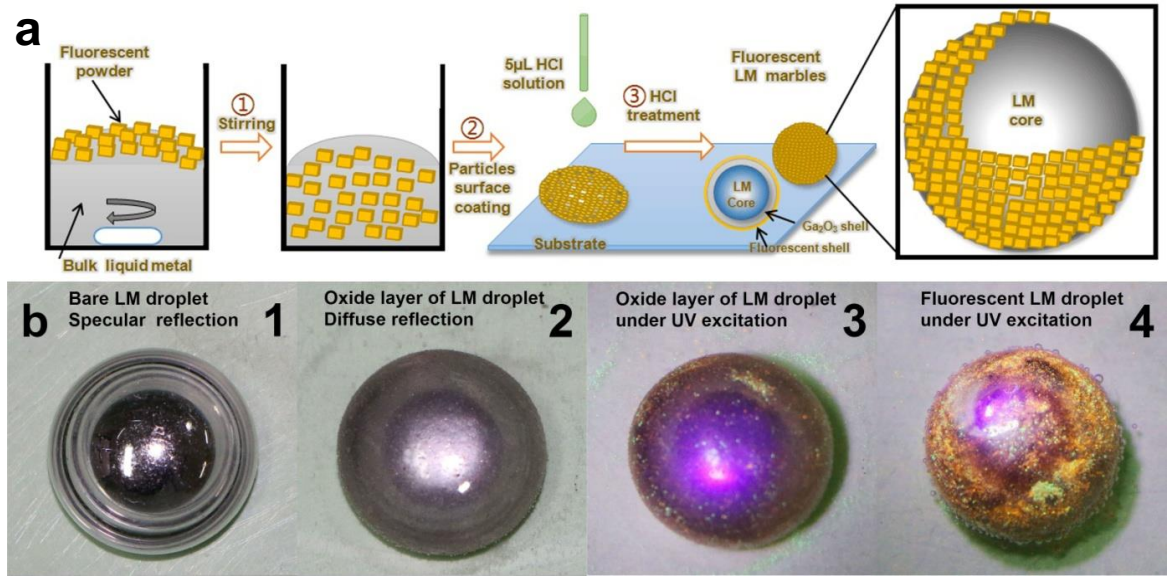

**Figure 1.** (a) Fluorescent LMs were prepared by stirring a LM marble with nano/micro-fluorescent powders in a beaker; (b) The color-changing of LM: (1) Bare EGaIn droplet; (2) Images of a EGaIn droplet with naturally formed native oxide layer in ambient air; (3) Images of EGaIn droplet with naturally formed native oxide skin under UV excitation; (4) The orange color performance of fluorescent LM was observed under UV excitation.

After being mixed with the fluorescent powders, the LM became more viscoelastic. It was amorphous and not easy to flow, like a solid. Take a spoonful of the mixture of fluorescent LM, then dealt it with hydrogen chloride (HCl) solution (6% wt.%). As an example: A fluorescent LM droplet of 10mL would be treated with 5μL HCl. After this treatment, the shape of the fluorescent LM was totally transformed into a "sphere" [46], and the physical properties of this marbles have changed from wetting to non-wetting state. The size of the LM droplets depends on the volume of the removal.

**Figure 1 (b-1)** shows a bare LM galinstan droplet (R=2.04 cm) with a shiny silver tone. The naturally oxidized LM droplet with a dull gray appearance was presented in **Figure 1 (b-2)** as well. Under UV excitation, this oxide skin was presented with a purple hue, as disclosed in **Figure 1 (b-3)**. And the fluorescent LM droplet with an orange appearance was given in **Figure 1 (b-4)**. The current discoveries presented a novel opportunity for the implementation of full-color appearances of LM.

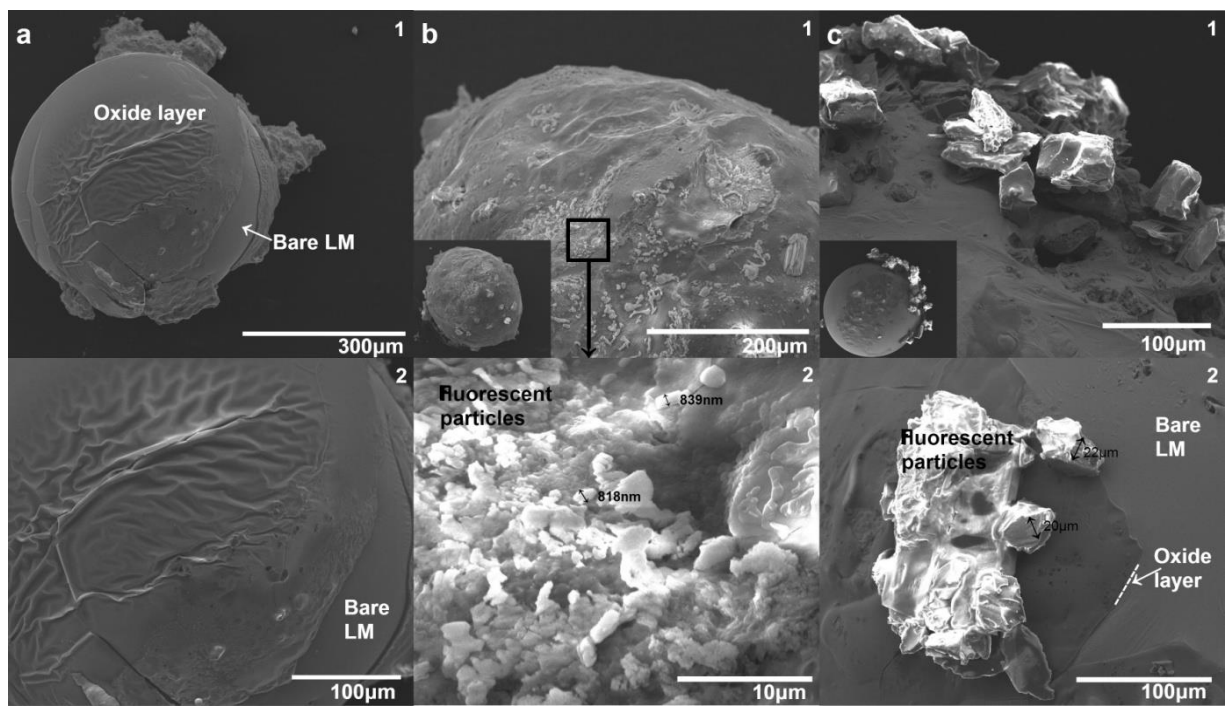

**Figure 2.** (a) (1)-(2) The SEM of oxidized LM with a draped and viscoelastic oxide layer. (b) The SEM images of a green fluorescent LM droplet: (1)-(2) LM droplet encapsulated in coating of 800 nm green fluorescent particles; (c) The SEM images of an orange fluorescent LM droplet: (1)-(2) LM droplet encapsulated in coating of 20 μm orange fluorescent particles.

In the following, the adhesion mechanism between the fluorescent particulates and LM marbles was investigated. **Figure 2 (a-1~2)** presented the images of the native oxide tegument which was naturally developed on the surface of a puny micro-fluorescent LM marble (R=480μm) in ambient air. The oxidized LM droplet has a draped and irregular oxide skin, and the layer was to protect the metal from further oxidation. The LM droplet was fast oxidized even in a mere 0.2% oxygen environment [45]. It has also been revealed that the "tearing morphology" was constructed from the expansion of the LM marble under vacuum in the SEM system [33]. The SEM images of some green and orange fluorescent LM marbles were presented in **Figure 2 (b)~(c)**. It can be observed that LM marbles stabilized with fluorescent particulates are multifaceted systems composed of LM cores and solid fluorescent particulates shell. The green and orange fluorescent particles have dimensions from 800 nm to 20 μm, respectively. The native oxide covering played a paramount role in adhesive bonding between the external functional particles and the bare LM marbles, as demonstrated in **Figure 2 (c-2)**. The results indicated that the fluorescent particles only existed on the surface of an oxide layer, and the particles could not adhere to the smooth and bare LM droplet.

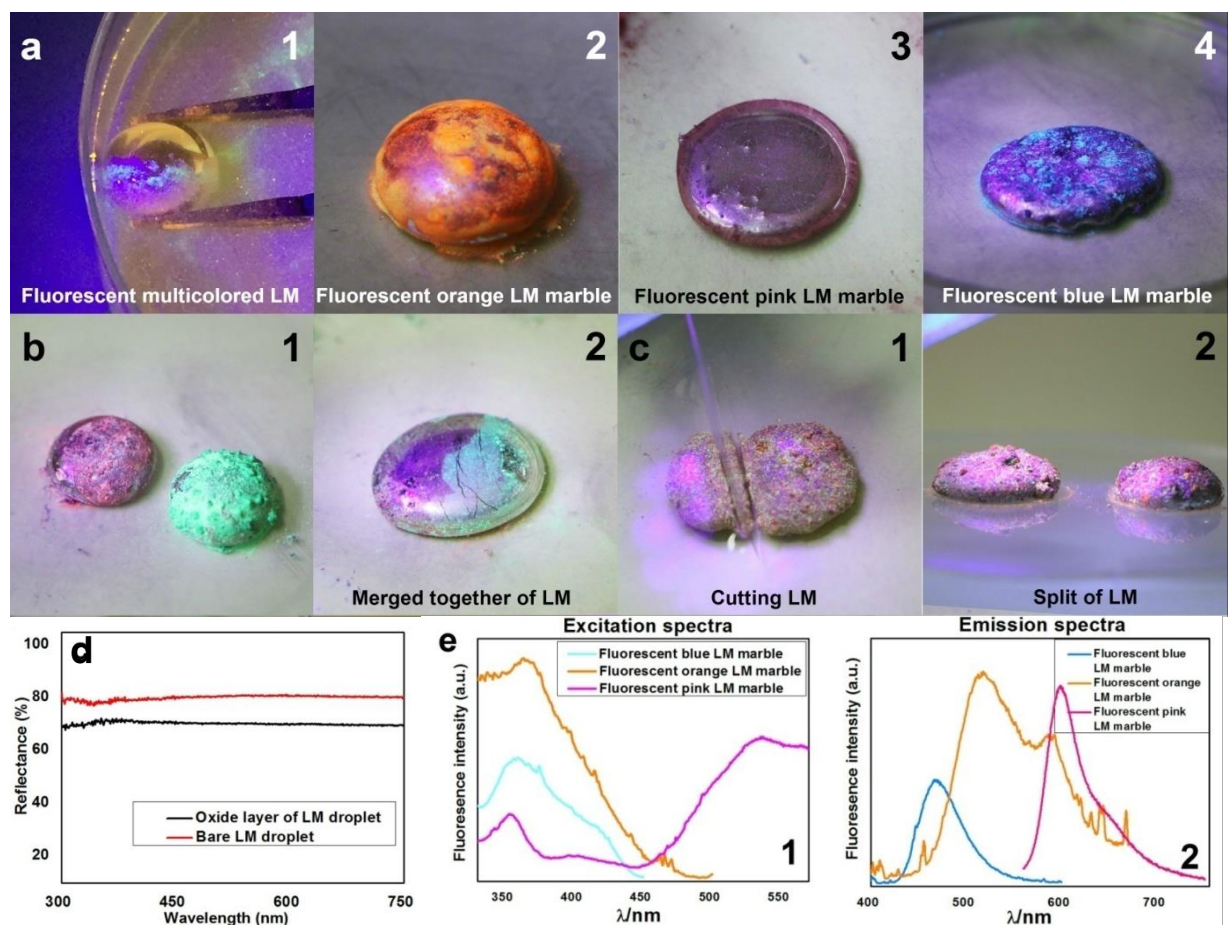

**Figure 3.** The fluorescent liquid metal marbles: (a) Luminescence of fluorescent LM marbles observed by the naked eye upon excitation with UV light from left to right: (1) Optical image of fluorescent multicolored LM marble (R=1.7cm), (2) Orange LM marble (R=2.3cm), (3) Pink LM marble (R=3.09cm), and (4) Blue LM marble (R=3.55cm). (b) The merged together color from different fluorescent LM marbles: (1) A fluorescent pink LM marble and a fluorescent green LM marble; (2) The composite fluorescent LM marble constructed from the pink and green LM marbles. (c) The split of fluorescent LM marble: (1) A fluorescent LM marble; (2) Cutting the fluorescent LM marble; (d) The NIR reflectance of bare LM and oxidized LM; (e) The excitation (1) and emission (2) spectra of three different fluorescent LM marbles in the figure (a 2-4).

Different synthesized fluorescent LM marbles were manufactured with either relatively small or large sizes, as clearly exhibited in **Figure 3 (a),** such as multicolored fluorescences, fluorescent orange, pink, to blue colors etc. The oxide film makes it much easier for LM to adhere to the solid surface and also stabilize LMs into different shapes. When constructed, these fluorescent LM marbles were "sphere" in shape, but occasionally adopted an "ellipsoidal" geometry due to their dissimilar surface tensions.

Besides, the addition of fluorescence has not changed the particular physical-chemistry properties of LM droplet, which has been explained in the previous paper [47]. **Figure 3 (b)** presented a photograph of the originally separated LM droplets with pink and green appearances. A straw was used to merge the two different colored LM droplets together, which resulted in a biggish ultimate multicolor LM marble. In **Figure 3 (c)**, utilizing a blade, it can divide one entire fluorescent LM marble into a couple of slight LM marbles. This cutting manipulation can be performed more than once and one huge fluorescent LM marble can be transformed into multitudinous miniature multicolor LM spheres. Consequently, the fluorescent LM marbles have the power to be split into tens of thousands of multicolor LM marbles or merged together between distinctive droplets.

The color of an object depends on its physics. Objects can display the color of the light leaving their surfaces, which normally depends on the reflectance properties of the surface. Therefore, by testing the Vis-reflectivity of different colorful LM marbles, the optical properties can be known. An excellent Vis reflectance (81%) of the bare LM marble with a specular reflection was exhibited in **Figure 3 (d)**, while the Vis reflectance of the oxidized LM marble with an oxide skin diminished to 72%. The oxidized LM has a lower reflectance of light, so its color is darker when compared with the bare LM. When different bands of light were used to excite the fluorescent LM, the fluorescent LM can emit different colors of fluorescence in the dark. As shown in **Figure 3 (e-1)**, the fluorescent orange, pink, and blue LM marbles displayed excitation peaks around 358nm, 364nm, and 535nm, respectively. The fluorescent emission spectrum of these marbles manifested narrow and symmetric emission peaks at 467nm, 517nm, and 600nm, respectively, as shown in **Figure 3 (e-2)**. Since the fluorescent particulates are self-luminous materials, the intense bright luminescence of dissimilar fluorescent LM marbles could be perceived by the naked eyes in the dark.

In the following, we spread a drop of 2cm diameter fluorescent LM on the surface of the plastic substrate and illuminated it with an UV light. The major phenomenon in the basic electrolytic solution (NaOH) and acidic electrolytic solution (HCl) (about 6% wt.%) were summarized in Table 1 and **Figure 4**, respectively.

When the fluorescent LM was contacted with the cathode (-) of copper electrode at the voltage of 15V, it was found that the fluorescent LM marble slowly released the fluorescent nano/micro-particles. The fluorescence firstly disappeared and the bare LM surface occurred near the cathode at 293 seconds. The LM changed appearance from its fluorescent orange color into a whole naked one at 380 seconds. This fluorescent LM was prepared from the fluorescent powder of 2g and the LM of 20g. Therefore, almost 9.09% (wt.%) phosphor could be completely released from LM in about 420 seconds. Typical snapshots of this kind of discoloration are presented in **Figure 4 (a)** and Table 1-case 1 and 4, respectively.

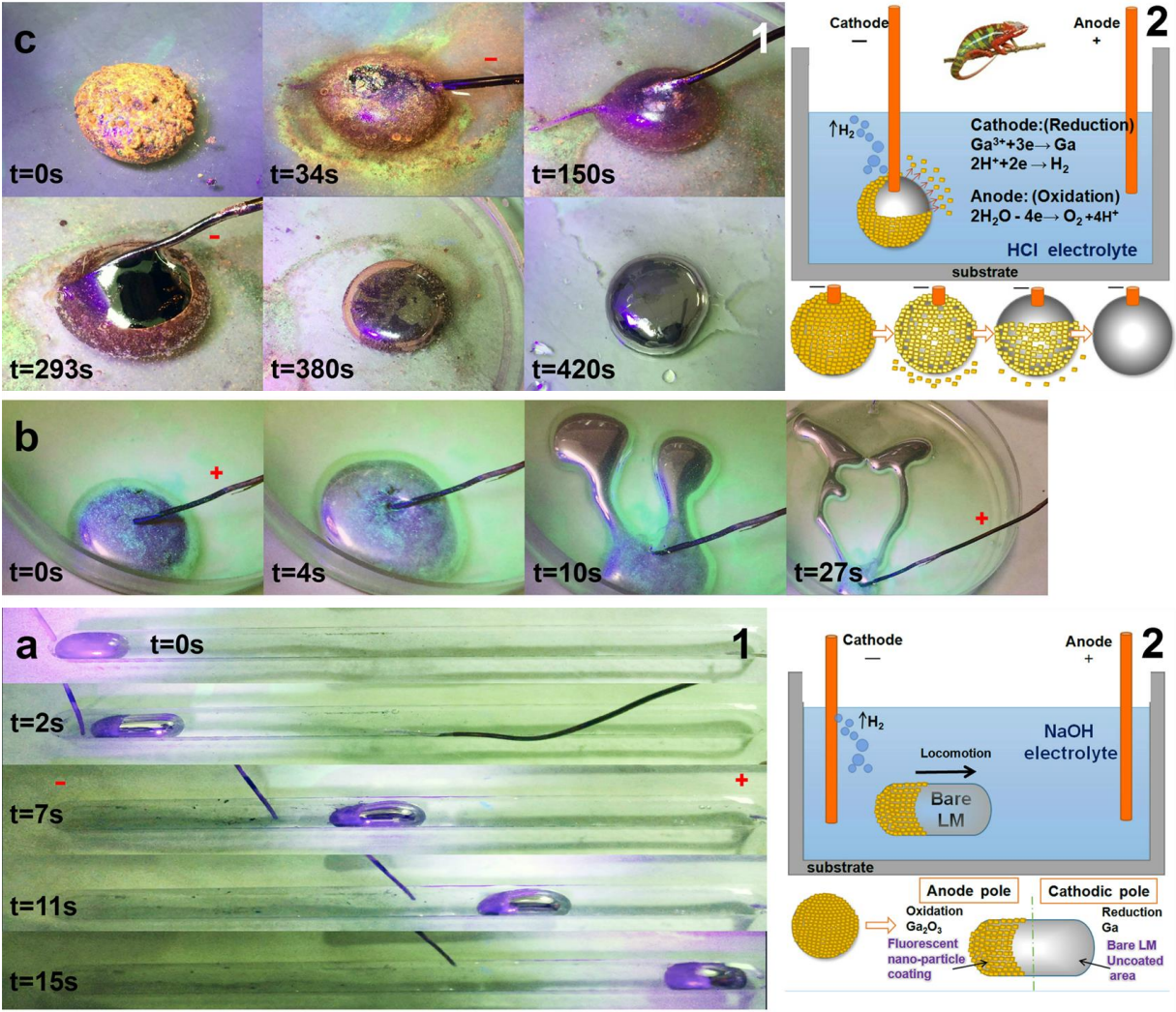

**Figure 4.** (a) Electrically induced discoloration of the fluorescent LM marbles in a HCl electrolyte: (1) Schematic of the electric field force influencing the discoloration of the fluorescent LM marble in a basic HCl solution; (2) The mechanism of the process to release the fluorescence. (b) Electrically induced transformation of the fluorescent LM marbles in a NaOH electrolyte; (c) Movement of the fluorescent LM marbles in a basic electrolyte when a step voltage of 15V is applied across the two copper electrodes: (1) Actuation of fluorescent LM marbles; (2) The mechanism of the process of the fluorescent LM marble in electric field.

When introducing electrical stimulation, the surface layer of $Ga_2O_3$ was reduced in the cathode, producing bare Ga and indicating significant variations of the adhesion property.

The reaction occurs at the cathode (-): $2H^+ + 2e \rightarrow H_2\uparrow$ (1) $Ga^{3+} + 3e \rightarrow 2Ga$ (2)

The reaction occurs at the anode (+):    $2H_2O - 4e \rightarrow O_2 + 4H^+$    (3)

As a consequence, the fluorescent particulates were constantly released from the oxide tegument. The discoloration was also examined and interpreted in NaOH and HCl solutions.

The loading of fluorescence on the LM droplet depends on the area of the epidermis of LM. The volume and surface area of the LM droplet is $4/3\pi r^3$, and $4\pi r^2$ (suppose the radius= r), respectively. And the oxide film and nano/micro phosphor were only 0.5~2.5nm and 0.8~20 μm thick, respectively. So the volume of the fluorescence reads as $\approx 4\pi r^2 \times 20\mu m$. Thus:

the loading of the fluorescent material = $(4\pi r^2 \times 20\mu m)/(4/3\pi r^3) = 60/r$    (4)

Then, suppose there was a LM droplet with a diameter of 2cm. The loading of fluorescence on the LM droplet is 60*2/20000= 0.6%. That means the loading phosphor on a drop of LM of 2cm is 0.6%.

When the anode (+) was attached to the fluorescent orange LM, the copper-wire cathode (-) was immersed in water, the colorful LM sphere quickly transformed into a flat film with a large area, in Table 1-case 2 and 5. **Figure 4 (b)** illustrates the deformation of fluorescent green LM in the basic electrolyte. In the first 4 seconds, the LM extended gradually; while in the 10 seconds, the green LM moved close to the cathode, and the oxidation and deformation were observed on the bare surface of LM, which has also been explained in the previous paper [5]. If the fluorescent LM was immersed in HCl, it would quickly become a black flat LM film. The processes were also presented in Table 1-case 2 and 4. We speculated that the LM chemically reacted with HCl and turned into $GaCl_3$, which was also demonstrated by former research [46]:

The reaction occurs at the cathode (-):    $2H_2O + 2e \rightarrow 2(OH)^- + H_2 \uparrow$    (5)

The reaction occurs at the anode (+):    $2(Cl)^- - 2e \rightarrow Cl_2 \uparrow$    (6)

$$2Ga_2O_3 + 6HCl \rightarrow 2GaCl_3 + 3H_2O$$    (7)

$$2Ga + 6HCl \rightarrow 2GaCl_3 + 3H_2$$    (8)

When the electrodes were immersed in the electrolyte without contacting the fluorescent LM, the fluorescent LM marble would motivate towards the anode at a speed of 0.5 body length per-second both in NaOH and HCl solutions, as displayed in **Figure 4 (c)** and **Table 1**-case 3 and 6, respectively. What is more, the fluorescent nano-particles on the surface of the LM marble were changed to non-uniform. The cathodic pole of LM was reduced, and a bare LM with the uncoated area was formed. The $Ga_2O_3$ integument full of fluorescent powders was created on the anodic pole hemisphere.

Through changing the electrode arrangements and the electrolytes, various behaviors of fluorescent LMs were investigated, including deformation, discoloration, and the planar locomotion, respectively. Clearly, using this electric-stimuli can manipulate the LM shapes and colors, which would help simulate the interestong behaviors of the chameleon in nature. Besides, this is also a first ever report for an electricity-responsive LM to achieve efficient releasing of fluorescent substances.

**Table 1 The phenomena of fluorescent LM marbles in different electrolytes**

| Electrolyte | In NaOH | | | Phenomena |
|---|---|---|---|---|
| Case | Anode (+) | Cathode (-) | (+)(-) | |
| 1 | Not contacted | Contact LM | | Discoloration |
| 2 | Contact LM | Not contacted | | Transformation |
| 3 | | | Not contacted | Planar locomotion |
| | In HCl | | | |
| | Anode (+) | Cathode (-) | (+)(-) | |
| 4 | Not contacted | Contact LM | | Discoloration |
| 5 | Contact LM | Not contacted | | Transformation, turn black |
| 6 | | | Not contacted | Planar locomotion |

## 4. Conclusion

In summary, we have demonstrated a facile and rapid route to successfully fabricate fluorescent liquid metal which could work as a transformable biomimetic chameleon. Such approach takes advantage of the inherent adhesion properties at the oxide surface of the LM marbles. The original liquid metal with a single silvery-white metal color can be manipulated into full-colors. And it still has unique physical/chemical, and optical properties, such as splitting or merging. We disclosed that this fluorescent LM are able to respond to a particular electric stimulus, and change their original shapes and colors just like a transformable biomimetic chameleon. Moreover, under the electric field stimulation, all these nano or micro materials in the outside layer of the LM can be completely released. Such electro-responsive materials are smarter and more intelligent than ordinary materials, and the present finding has both fundamental and practical significance. The fluorescent liquid metal possesses unique virtues in a wide variety of crucial areas, such as functional nanocomposites, controlled/targeted matter delivery, and color changeable bionic soft robot in the coming time.

## Acknowledgements

This work is partially supported by the Ministry of Higher Education Equipment Development Fund, Dean's Research Funding and the Frontier Project of the Chinese

Academy of Sciences, as well as Beijing Municipal Science & Technology Funding (Grant No. Z151100003715002).